\documentclass{aa}
\usepackage{array}
\usepackage{amsmath}
\usepackage{amssymb}
\usepackage{graphics}
\usepackage{natbib}
\bibpunct{(}{)}{;}{a}{}{,} 
\defcitealias{Hartiganetal95}{HEG95}
\defcitealias{Garcia2001a}{Paper~I}
\defcitealias{Lavalley2000}{LF2000}

\begin{document}

%
   \title{Atomic T~Tauri disk winds heated by ambipolar diffusion}
   \subtitle{II- Observational tests}

   \author{P.~J.~V.  Garcia\inst{1,2,}\thanks{CAUP Support Astronomer,
              during  year   2000,  at  the  Isaac   Newton  Group  of
              Telescopes, Sta.  Cruz de La Palma, Spain},
           S. Cabrit\inst{3},
           J. Ferreira\inst{4},
          \and
           L. Binette\inst{5}}

   \offprints{P.~J.~V. Garcia \email{pgarcia@astro.up.pt} }

   \institute{Centro de Astrof\'{\i}sica da Universidade do Porto,
              Rua das Estrelas, 4150-762 Porto, Portugal
              \and
              CRAL/Observatoire de Lyon, CNRS UMR 5574, 9 avenue
              Charles Andr\'e, 69561 St. Genis-Laval Cedex, France
              \and
              Observatoire de Paris,
              DEMIRM, UMR 8540 du CNRS,
              61 avenue de l'Observatoire,
              F-75014 Paris, France
              \and
              Laboratoire d'Astrophysique de l'Observatoire de Grenoble,
              BP 53, 38041 Grenoble Cedex,  France
              \and
              Instituto de Astronom\'\i a,
              UNAM, Ap. 70-264, 04510 D. F., M\'exico
}

   \date{Received ??/ Accepted ??}

   \authorrunning{Garcia, P., et al.}

   \titlerunning{Atomic T~Tauri disk winds heated by ambipolar diffusion}

   \abstract{
     Thermal    and    ionization    structures    of    self-similar,
     magnetically-driven  disk  winds obtained  in  a companion  paper
     \cite{Garcia2001a} are used to  compute a series of jet synthetic
     observations.   These include  spatially resolved  forbidden line
     emission maps,  long-slit spectra, as  well as line  ratios. Line
     profiles and  jet widths  appear to be  good tracers of  the wind
     dynamics and  collimation, whereas line  ratios essentially trace
     gas  excitation  conditions.\\  
     All  the  above diagnostics  are  confronted  to observations  of
     T~Tauri  star microjets.   Convolution by  the observing  beam is
     shown to  be essential for a  meaningful test of  the models.  We
     find that jet widths  and qualitative variations in line profiles
     with  both distance  and  line tracers  are  well reproduced.   A
     low-velocity [\ion{O}{i}] component is also obtained, originating
     from the  disk wind base.   However, this component is  too weak,
     predicted  maximum velocities  are  too high  and electronic  and
     total densities are too  low. Denser and slower magnetized winds,
     launched from disks with  warm chromospheres, might resolve these
     discrepancies.
}

   \maketitle

%

\section{Introduction}

Jets  from   young  protostars  such   as  T~Tauri  stars   remain  an
enigma. Indeed,  despite tremendous efforts,  both observationally and
theoretically,  we do not  know whether  jets are  driven by  the disk
alone  (so  called  ``disk   winds''),  the  central  protostar  alone
(``stellar  winds'') or the  interaction zone  between the  disk inner
edge and  the protostellar  magnetosphere (``X-winds''). As  argued in
\cite[  hereafter Paper~I]{Garcia2001a}  probing  the ultimate  energy
source   powering  T~Tauri   jets   requires  reliable   observational
predictions from any  dynamical jet model.  To do  so, the thermal and
ionization states of the gas must  then be solved along the flow. This
remains to be  addressed in a fully self-consistent  way, leaving room
for speculations.

In  contrast   with  recent   approaches  which  assumed   a  constant
temperature   and  ionization  profile   \citep{Shang98,Cabrit99},  in
\citetalias{Garcia2001a} we solved {\em  a posteriori} the thermal and
ionization  structures of  a magnetically-driven  disk wind  heated by
ambipolar diffusion.  Indeed, of  all possible heating mechanisms only
ambipolar diffusion  heating allows ``minimal''  thermal solutions, in
the sense  that the same  physical process ---  non-vanishing currents
--- is  responsible   for  both  jet  dynamics  and   heating.   As  a
consequence  no additional tunable  parameter is  invoked for  the jet
thermal  description.  We  therefore extended  the pioneering  work of
\cite{Safier93a,  Safier93b} by  (1) using  magnetically-driven (cold)
jet   solutions,  self-consistently   computed  with   the  underlying
accretion  disk \cite{Ferreira97};  (2) a  more accurate  treatment of
ionization using  the {\tt Mappings Ic} code  and ion-neutral momentum
exchange rates including the thermal contribution.

As in  \cite{Safier93a}, we obtained  jets with a  temperature plateau
around $10^4$~K, but ionization fractions revised downward by a factor
of 10-100. This  last fact is due to the  previous omission of thermal
speeds in ion-neutral momentum-exchange rates and to our different jet
solutions.   Moreover,  the physical  origin  of  the hot  temperature
plateau   has  been   shown  to   represent  a   robust   property  of
magnetically-driven disk winds heated by ambipolar diffusion. Last but
not least, we showed that the calculated thermal and ionization states
remain consistent  with the  approximations used for  jet calculations
\citepalias[see   Appendix~C,][]{Garcia2001a}:    (1)   single   fluid
description of a (2) thermalized  (all chemical species share the same
temperature), (3) cold (negligible  thermal pressure gradient) and (4)
perfectly conducting  plasma. We  stress that such  consistency checks
are of an extreme physical importance and must therefore be made.

In this paper, we confront the detailed model calculations obtained in
\citetalias{Garcia2001a} with observations  of forbidden line emission
regions  in T~Tauri  stars, believed  to probe  accretion-driven winds
\citep[][ hereafter  HEG95]{Cohen89,  Cabrit90,  Hartiganetal95}.   The
high-velocity emission  has been spatially resolved in  a dozen stars,
and found to arise from  collimated ``microjets'' extended over only a
few 100~AU  \citep[][ and refs.   therein]{Solf89, Kepner93, Hirth94a,
Hirth94b,   Burrows96,  Ray96,  Hirth97,   Bacciotti99,  Lavalley2000,
Dougados2000a,  Bacciotti2000a}.  These microjets  represent currently
our best opportunity to test  accretion-driven jet models: (1) At such
small  distances  from  the  star,  the  intrinsic  jet  structure  is
(hopefully) less  perturbed by instabilities or  interactions with the
ambient medium  than in  the more distant  jet sections  accessible in
infrared sources; (2) Direct, albeit uncertain, estimates for the disk
accretion  rate have  become recently  available from  measurements of
continuum    excess    in    the   stellar    photospheric    spectrum
(\citetalias{Hartiganetal95}, \citealt{Gullbring98}).

We  point out that  the ``correct''  jet model  should pass  {\em all}
available observational tests, namely:  microjet widths as measured by
adaptive optics and the Hubble Space Telescope, long slit spectra, and
integrated line profiles, line fluxes and line ratios.  Note that only
the latter constraints based  on integrated profiles were available to
\cite{Safier93b}. We  are now in  a position to perform  more detailed
tests  of the  class of  self-similar, magnetically  driven  cold disk
winds, involving spatially-resolved  diagnostics.   We also stress
that given the steep gradients  in physical conditions at the base of
the  wind,  meaningful comparison  with  observations requires  proper
convolution of maps and long-slit spectra by the observing beam.

This article  is structured  as follows. Section~2  is devoted  to the
predicted  forbidden line  emission diagnostics  and  their comparison
with  T~Tauri  stars  observations.    We  conclude  in  Section~3  by
summarizing the successes and failures of our model, and by discussing
perspectives to resolve the discrepancies.

\section{Predicted forbidden line emission and comparison with T Tauri
microjets} 

We  used  three  different  models  of  self-similar  magnetized
accretion-ejection structures \citep[MAES,][]{Ferreira97}, labelled A,
B and C. They are mainly characterized by a different ejection index
\begin{equation}
\xi = \frac{d \ln \dot{M}_{\text{acc}}(\varpi) }{d \ln \varpi}
\end{equation}
measuring the local  ejection efficiency, where $\dot{M}_{\text{acc}}$
is the  disk accretion rate  at the cylindrical radius  $\varpi$. Note
that $\xi=0$ in a standard accretion disk. In our models, the ejection
index controls the jet opening, a smaller $\xi$ providing a larger jet
radius. Model  A is obtained  for $\xi=0.01$, model B  for $\xi=0.007$
and  model   C  for  $\xi=0.005$.    A  magnetized  accretion-ejection
structure is  assumed to  be settled around  a protostar  of $M_*=0.5\
{\rm M}_\odot$, between  an inner and an outer  edge of, respectively,
$\varpi_{\rm  i}=0.07 {\rm  AU}$  and $\varpi_{\rm  e}=  1 {\rm  AU}$.
Outside this  radius, we assume  the presence of a  standard accretion
disk  fueling the  MAES with  mass (and  possibly magnetic  flux). The
accretion rate  is taken to  vary between $10^{-8}$ and  $10^{-5} {\rm
M}_\odot  {\rm  yr}^{-1}$.   We  refer  the  reader  to  Section~2  in
\citetalias{Garcia2001a} for further  details concerning the dynamical
models and for a discussion of our choice of parameters.

\subsection{Calculation of synthetic emission maps and profiles}

Synthetic emission maps and long  slit spectra along the jet axis were
computed   in  the   lines  of   [\ion{O}{i}]6300,  [\ion{S}{ii}]6731,
H$\alpha$, and  [\ion{N}{ii}]6584, using the  emissivities provided by
{\tt Mappings  Ic}. Emission  was assumed optically  thin, simplifying
calculations  of the  emergent profiles.   This assumption  is clearly
adequate for forbidden lines,  though possibly not for H$\alpha$ whose
opacity is probably dominated by scattering of Ly$\alpha$ photons. Our
results for the latter line  are therefore only indicative, a complete
treatment of  the line radiative  transfer being outside the  scope of
the present  paper.  Therefore only  forbidden lines will be  used for
comparison  with observations  of T~Tauri  stars, where  the H$\alpha$
line  is contaminated  by a  strong magnetospheric  contribution which
will probably further affect the radiative transfer.

In order to  adequately sample the innermost jet  regions where strong
gradients in density  and velocity are present, we  used a logarithmic
sampling  in $x  = z/\varpi$  and in  $\varpi_0$  between $\varpi_{\rm
i}=0.07 {\rm AU}$ and $\varpi_{\rm  e}= 1 {\rm AU}$ and projected each
cell onto  a rectangular  grid in  the plane of  the sky  with 0.25~AU
pixel size.   Maps were  convolved with a  28~AU (FWHM)  gaussian beam
corresponding  to 0.2\arcsec\ at  the distance  of Taurus,  typical of
resolutions  accessible  with  adaptive  optics or  the  Hubble  Space
Telescope.

To construct  long slit spectra,  line emission from each  cell volume
was  distributed  in radial  velocity  bins  of  3.5~km/s, and  summed
perpendicular  to the jet  axis, keeping  only the  projected distance
from the  star, $d$, as  spatial coordinate. The result  was convolved
with  a 70~AU  (FWHM) gaussian  along $d$  and with  a  10~km/s (FWHM)
gaussian along the velocity axis, corresponding to typical resolutions
achieved in ground-based  long-slit spectra of microjets \citep[e.g.][
hereafter LF2000]{Solf93, Lavalley2000}.

Note  that qualitative  forbidden line  predictions for  model  A were
presented by \cite{Cabrit99} using a highly simplified temperature and
ionization prescription  for the  plateau.  Comparison of  the present
accurate results with this preliminary  work will be used to probe the
sensitivity of  maps and line  profiles on the underlying  thermal and
ionization structure.

\subsection{Wind morphology: Images and jet widths}

\begin{figure}
\begin{center}
\resizebox{8cm}{!}{\rotatebox{-90}{\includegraphics{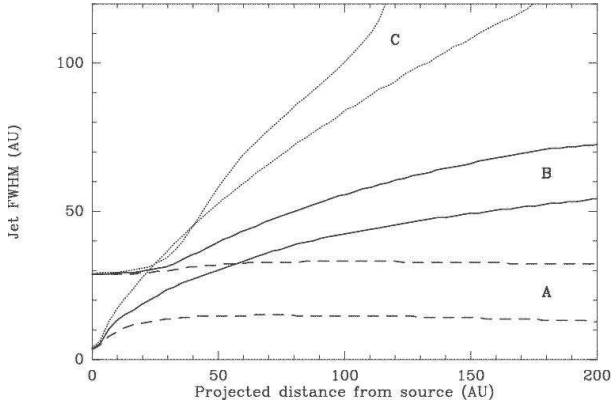}}}
\end{center}
\caption{Predicted jet FWHM (gaussian fit) as a function of distance
to the source for  our models A, B, C (With $\xi  = 0.01,\ 0.007\ {\rm
and}\  0.005$)  and  two  different  beam  sizes:  28~AU  and  2.8~AU,
corresponding   in  Taurus  to   0.2\arcsec\  and   0.02\arcsec\  (VLT
diffraction  limit).  Note the  strong  dependence  on  $\xi$ and  the
considerable bias introduced by the beam size.}
\label{fig:fwhm} %
\end{figure}
  
\begin{figure*}
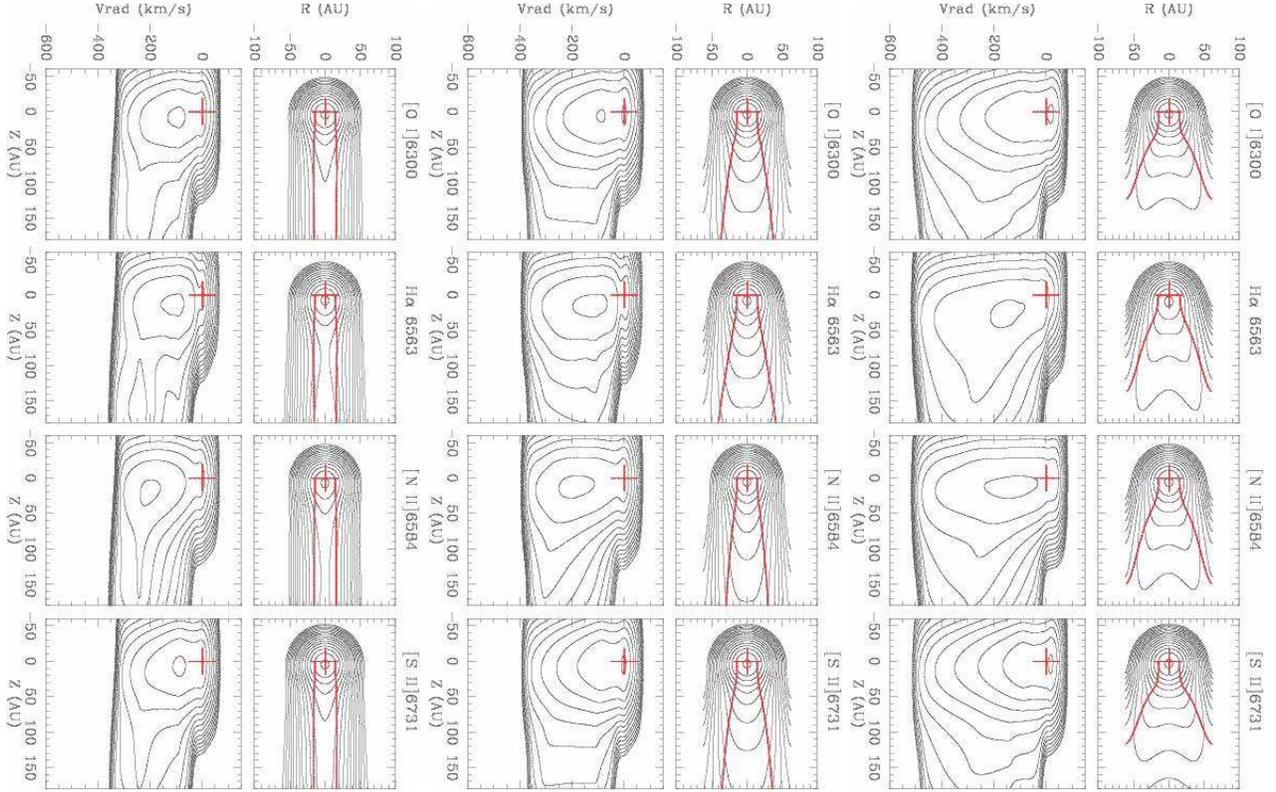

\begin{center}
\resizebox{5.5cm}{!}{\rotatebox{-180}{\includegraphics{fig9a.epsi}}}
\resizebox{5.5cm}{!}{\rotatebox{-180}{\includegraphics{fig9b.epsi}}}
\resizebox{5.5cm}{!}{\rotatebox{-180}{\includegraphics{fig9c.epsi}}}
\end{center}
\caption{Predicted maps and long-slit spectra in various emission lines
for Models  A, B, C  ($\xi = 0.01,\  0.007\ {\rm and}\ 0.005$  left to
right).  Intensity  maps are  convolved with a  28~AU beam,  and thick
grey lines plot  the jet transverse FWHM (gaussian  fit) as a function
of position.  Long-slit spectra along  the jet axis are convolved by a
$70\  {\rm   AU}  \times  10\  {\rm  km/s}$   point  spread  function,
representative  of current ground-based  spectro-imaging performances.
Inclination      is       $i=60\degr$      from      pole-on,      and
$\dot{M}_{\text{acc}}=10^{-6}\,  {\rm  M}_{\odot} \text{yr}^{-1}$.   A
cross  marks the  stellar  continuum position.   Contours decrease  by
factors of 2.}
\label{fig:tout_82} %
\end{figure*}

Synthetic emission  maps are presented in  Fig.  \ref{fig:tout_82} for
our 3  models and typical parameters. They  successfully reproduce two
key  features observed  in microjets:  (1) a  strong,  unresolved peak
slightly  shifted  from the  stellar  position,  (2)  a jet-like  body
appearing  collimated within  200~AU of  the star.   As in  the X-wind
model  of  \cite{Shu94a}, this  jet-like  appearance  is an  ``optical
illusion''  reflecting  the collimation  of  the innermost,  brightest
streamlines; outer streamlines extend over a wider angle but have much
lower  surface brightness  \citep[cf.   Fig.~1 in][]{Cabrit99}.   Note
that the  MHD disk wind  solutions considered by  \cite{Safier93a} did
not collimate over such scales.

The shift of  the unresolved peak from the star  depends little on the
value of  $\xi$ or the chosen  line, but decreases  slightly for lower
$\dot{M}_{\rm  acc}$,  as  heating   then  occurs  sooner  along  each
streamline  (see  Sect.~4,  Paper~I).   It  is $\sim$  5~AU  at  28~AU
resolution for $\dot{M}_{\rm  acc}$ = $10^{-6}$ M$_{\odot}$ yr$^{-1}$,
smaller  than observed  \citep[15-30~AU][]{Solf93,  Hirth94a, Hirth97,
Lavalley97}.  However, this does not  represent a reliable test of the
model  as observed  shifts may  be  severely biased  by inaccuracy  in
continuum  subtraction  \citep[a 2\%  error  in  subtraction causes  a
$\simeq  4\  {\rm AU}$  error  for  bright  jets;][]{Garcia99} and  by
possible obscuration by a flared disk.

In  contrast, Figures~1  and  2 show  that  the jet  width provides  a
powerful test of the models: It depends very little on the chosen line
tracer,  inclination, jet  outer radius,  and mass-accretion  rate (at
least within our  explored range). It is also  not overly sensitive to
the detailed  ionization and  temperature structure, since  we recover
very   similar   widths  as   in   the   simplified  calculations   of
\cite{Cabrit99},  for the  same  wind solutions.   For  this class  of
models, the  jet FWHM depends mostly  on the degree of  opening of the
innermost          streamline,         fixed          by         $\xi$
(Figs.~\ref{fig:tout_82},\ref{fig:fwhm}).    Note  that  it   is  also
severely biased by the  beam size (Fig.~\ref{fig:fwhm}), an example of
the importance of proper convolution for comparison with observations.
 
Measurements  of   microjet  widths  at   0.1\arcsec\  --  0.2\arcsec\
resolution  were   recently  presented  by   \citep{Burrows96,  Ray96,
Dougados2000a}.  Jets  with no  clear contamination by  bowshock wings
give  strikingly consistent  results,  namely a  typical  jet FWHM  of
30-50~AU at  a distance of 200~AU.   Models A and B,  with $\xi \simeq
0.007\  \text {--}\  0.01$, reproduce  well these  observations, while
models with  smaller $\xi  < 0.007$, in  particular our model  C, seem
ruled out.

\subsection{Wind kinematics: Long-slit spectra and line profiles}

\begin{figure}
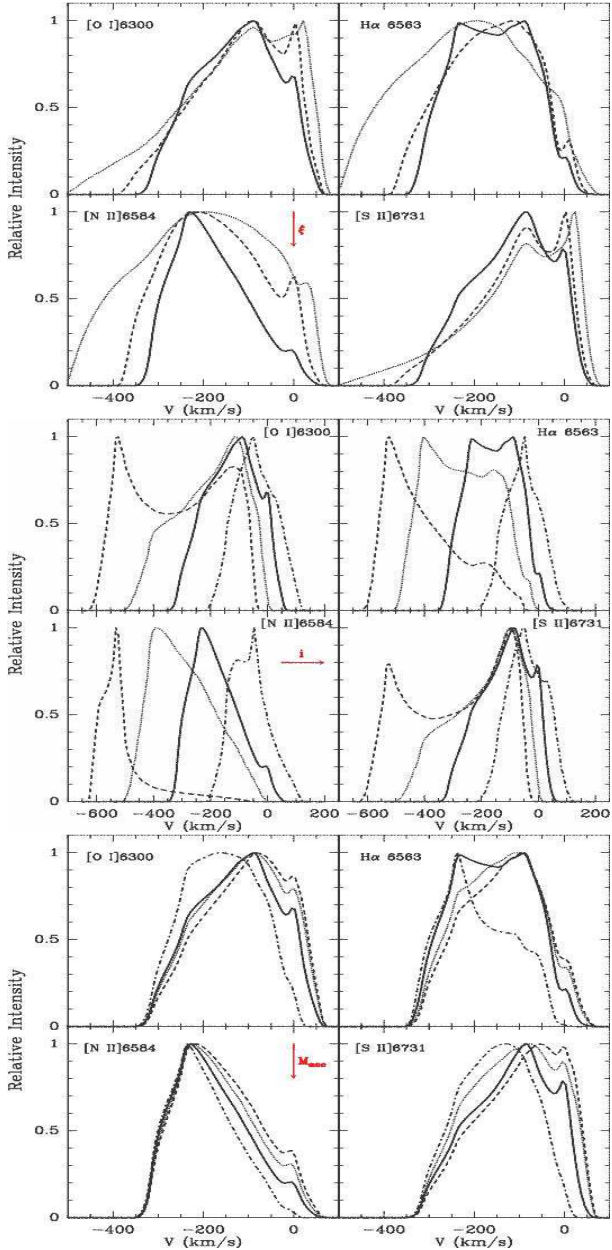

\begin{center}
        \resizebox{8cm}{5.5cm}{\includegraphics{4sol_new.epsi}}
        \resizebox{8cm}{5.5cm}{\includegraphics{4ang_new.epsi}}
        \resizebox{8cm}{5.5cm}{\includegraphics{4mdot_new.epsi}}
\end{center}
\caption{Integrated line profiles behavior.
{\bf  Top:}  Effect of  ejection  index  $\xi$:  Model A  ($\xi=0.01$,
solid),   B  ($\xi=0.007$,  dashed)   and  C   ($\xi=0.005$,  dotted).
Inclination      is      $i=60\degr$      and      accretion      rate
$\dot{M}_{\text{acc}}=10^{-6}\,  {\rm M}_{\odot}\text{yr}^{-1}$.  {\bf
Middle:}   effect    of   inclination   angle   for    Model   A   and
$\dot{M}_{\text{acc}}=10^{-6}$   M$_{\odot}$   yr$^{-1}$:  $i=20\degr$
(dashed),  $i=40\degr$ (dotted),  $i=60\degr$ (solid)  and $i=80\degr$
(dash-dot). {\bf  Bottom:} Effect  of accretion rate  for Model  A and
inclination    $i=60\degr$:    $\dot{M}_{\text{acc}}=10^{-8}\,    {\rm
M}_{\odot}   \text{yr}^{-1}$  (dashed),  $10^{-7}\,   {\rm  M}_{\odot}
\text{yr}^{-1}$  (dotted), $10^{-6}\, {\rm  M}_{\odot} \text{yr}^{-1}$
(solid) and $10^{-5}\, {\rm M}_{\odot} \text{yr}^{-1}$(dash-dotted)}
\label{fig:lines} %
\end{figure}

Long-slit   spectra    for   our    3   models   are    presented   in
Fig.~\ref{fig:tout_82} for  an inclination $60\degr$  from pole-on and
$\dot{M}_{\rm    acc}$    =     $10^{-6}    {\rm    M}_{\odot}    {\rm
yr}^{-1}$. Integrated line profiles  for a wider range of inclinations
and  $\dot{M}_{\rm acc}$  are presented  in  Fig.~\ref{fig:lines}. The
predicted profiles  successfully reproduce several  observed trends in
TTS:

- Two velocity components are  present: a low-velocity component (LVC)
formed  in very dense  regions at  the wind  base, just  where heating
starts, and a high-velocity component  (HVC), formed further up in the
accelerated wind regions.

- The  LVC  is stronger  in  [\ion{O}{i}]  than  in [\ion{N}{ii}],  as
observed   (\citetalias{Hartiganetal95},  \citealt{Hirth97}):   It  is
dominated by slowly rotating  streamlines with low ionization fraction
at $\varpi_0 \simeq$ 1~AU.

- The  relative  intensity  of  LVC  versus HVC  increases  for  lower
$\dot{M}_{\rm     acc}$     (Fig.~\ref{fig:lines}),    as     observed
\citep{Hartiganetal95}: This occurs because the hot plateau is reached
sooner for  lower accretion rates,  increasing emission from  the wind
base and hence the LVC.

- The HVC  is displaced  further from  the star than  the LVC  and the
displacement   is  larger  in   [\ion{N}{ii}]  than   in  [\ion{O}{i}]
(Fig.~\ref{fig:tout_82}),    as    observed   \citep{Hirth97}.     The
[\ion{N}{ii}] profile peaks at the  blue edge of the [\ion{O}{i}] line
profile       (Fig.~\ref{fig:lines}),      as       also      observed
\citep{Hartiganetal95,Hirth97}.  This  difference stems from  the slow
increase in ionization  fraction with distance \citepalias[cf.  Fig.~2
of][]{Garcia2001a}, which favors \ion{N}{ii} emission from the distant
accelerated wind regions.

- Long-slit spectra  show apparent acceleration  of the HVC,  with the
line profile peaking  around $-$100~km/s close to the  star and around
$-$300~km/s  at 150~AU.   Similar gradients  appear to  be  present in
\object{DG~Tau}   and   \object{UY~Aur}   \citep{Solf93,   Lavalley97,
Hirth97}.

The predicted profiles fail, however, in two respects:

- The relative intensity of LVC  versus HVC in [\ion{O}{i}] is smaller
than observed. The problem appears less severe in models with low $\xi
\le 0.005$ (see Model  C in Fig.~\ref{fig:lines}), as wider streamline
expansion decreases  the density in  the outer wind and  the resulting
HVC flux.  However,  such models predict jet widths  that are too wide
(see Fig.~\ref{fig:fwhm} and Sect.~2.2 above). The weakness of the LVC
in our  models reflect the  low temperatures encountered over  most of
the wind  base, where  low-velocity gas is  located. Note that  in the
simplified calculations of \cite{Cabrit99}, which assumed $T = 10^4$~K
above $z/\varpi \simeq 0.7$,  the LVC was more prominent, illustrating
the crucial influence of temperature at the wind base on the resulting
LVC intensity.

- Predicted maximum blue radial velocities in line profiles agree well
with     observations      of     the     \object{DG~Tau}     microjet
\citep[e.g.][]{Lavalley2000, Cabrit99}.  However  they appear too high
compared with  typical T~Tauri stars,  where $|V_{\rm blue}|  \le 200\
{\rm  km/s}$ \citep{Hartiganetal95,  Hirth97}, unless  most  stars are
observed    at    inclinations     $\ge    80\degr$    from    pole-on
(Fig.~\ref{fig:lines}).  Terminal velocities in  the present  class of
cold disk  wind models are  $\simeq V_{\rm kep}(\varpi_0)/\sqrt{\xi}$,
so even  our highest $\xi  = 0.01$ (model  A) tends to  give excessive
speeds.

\subsection{Integrated line fluxes}

\begin{figure}
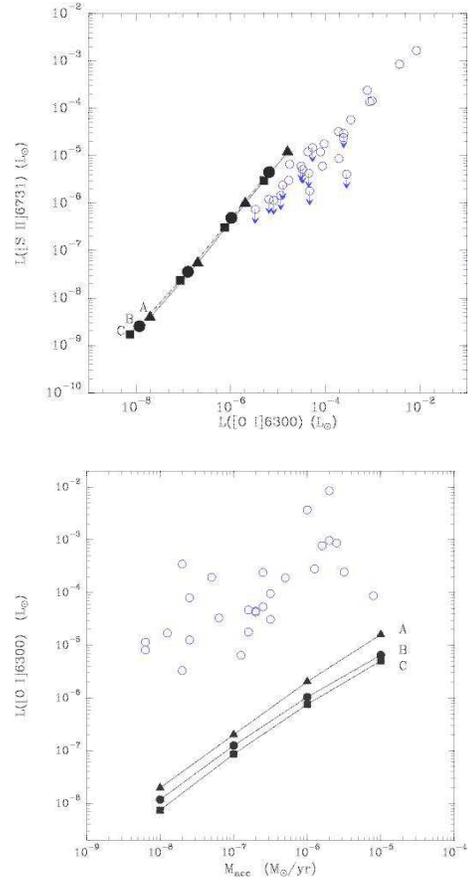

\begin{center}
\resizebox{6cm}{!}{\includegraphics{fig12a.epsi}}\\
\vspace{5mm}\resizebox{6cm}{!}{\includegraphics{fig12b.epsi}}
\end{center}
\caption{Output fluxes for model  A (triangles),
model B (circles) and model  C (squares). Model fluxes were integrated
out to 200~AU from the  star.  {\bf Top:} Integrated luminosity in the
[\ion{S}{ii}]6731  line plotted against  integrated luminosity  in the
[\ion{O}{i}]6300 line. Accretion  rates $\dot{M}_{\rm acc}$ range from
$10^{-8}$ to $10^{-5} {\rm M}_{\odot}  {\rm yr}^{-1}$ by factors of 10
(in the  direction of increasing  luminosity).  Observed T  Tauri star
forbidden  line luminosities from  \cite{Hartiganetal95} are  shown as
open circles, the arrows indicate upper limits on [\ion{S}{ii}].  {\bf
Bottom:} Integrated luminosities  in the [\ion{O}{i}]6300 line plotted
versus  $\dot{M}_{\rm  acc}$ for  our  models.   T  Tauri stars  total
[\ion{O}{i}] luminosities are shown as circles \citep{Hartiganetal95};
only  stars with detected  emission are  presented.  Models  predict a
roughly linear  law L[\ion{O}{i}]/$L_\odot\simeq\dot{M}_{\rm acc}/{\rm
M}_{\odot}{\rm  yr}^{-1}$, and  lie  typically a  factor $\simeq$  100
below observations.}
\label{fig:flux} 
\end{figure}

Figure ~\ref{fig:flux} plots  the predicted integrated luminosities in
the [\ion{O}{i}]$\lambda$6300 and [\ion{S}{ii}]$\lambda$6731 lines for
our 3 models and our whole range of $\dot{M}_{\rm acc}$ from $10^{-8}$
to  $10^{-5}$ M$_{\odot}$  yr$^{-1}$.  For  comparison,  observed line
luminosities in  T~Tauri stars with  {\it the same range  in accretion
rates}, taken from \cite{Hartiganetal95}, are also plotted.

Two characteristics  of the observations  are well reproduced:  (1) We
find   a   correlation    between   [\ion{O}{i}]   and   [\ion{S}{ii}]
luminosities,   with  a   slope  close   to  that   observed   in  TTS
(Fig.~\ref{fig:flux}a). (2)  The [\ion{O}{i}] line  luminosity is seen
to increase linearly with the accretion rate (Fig.~\ref{fig:flux}b), a
proportionality also observed in TTS \citep[cf.][]{Hartiganetal95}.

The  latter  proportionality  is  an interesting  consequence  of  the
opposite dependence of  $n_{\rm e}$ and $T$ on  $\dot{M}_{\rm acc}$ in
the hot  plateau: As $\dot{M}_{\rm  acc}$ increases from  $10^{-8}$ to
$10^{-5}$ M$_{\odot}$ yr$^{-1}$, $n_e$ increases effectively only by a
factor  of 10  (recall that  ionization fraction  decreases  almost as
1/$\dot{M}_{\rm acc}$),  while the plateau temperature  decreases by a
factor of a few  \citepalias[see Fig.~2 of][]{Garcia2001a}.  Given the
strong temperature  dependence of [\ion{O}{i}]  collisional excitation
rates  $q_{lu}$,  the  two  effects  cancel  out  and  the  emissivity
($\propto  q_{lu} n_{\rm  e}  n_{\rm O}$  in  the low-density  regime)
becomes   proportional  to   the   oxygen  density   only,  i.e.    to
$\dot{M}_{\rm acc}$.

Quantitatively, however,  the predicted fluxes  fail in that  they are
systematically too weak  compared to observations of TTS,  by a factor
100  on average.   We  find L([\ion{O}{i}])/$L_\odot\simeq\dot{M}_{\rm
acc}$/M$_{\odot}$  yr$^{-1}$.  In  this  respect, we  come  to a  very
different  conclusion from  \cite{Safier93b}, who  concluded  that his
model was able  to reproduce the range of  forbidden line luminosities
observed in  TTS.  There  are two main  reasons for  this discrepancy:
First,  our  lower rate  coefficients  for  H-H$^+$ momentum  exchange
result in a factor of 10 lower ionization fractions, $n_e$ values, and
hence  [\ion{O}{i}]  luminosities,   compared  to  Safier  (Paper  I).
Second, Safier considered mass-loss  rates ranging up to $10^{-6} {\rm
M}_{\odot}  {\rm  yr}^{-1}$, corresponding  to  a mass-accretion  rate
$\simeq  10^{-4} {\rm  M}_{\odot} {\rm  yr}^{-1}$, i.e.   an  order of
magnitude above  the maximum $\dot{M}_{\rm  acc}$ values in  TTS later
deduced by \citetalias{Hartiganetal95} from veiling measurements.

In the next section, we show that the flux deficit in our models stems
at least  in part from a lower  $n_{\rm e}$ than observed  in TTS, but
possibly also from an insufficient total density in our models.

\subsection{Forbidden line ratios and total densities}

The  ratios  of  the  [\ion{O}{i}], [\ion{S}{ii}],  and  [\ion{N}{ii}]
forbidden lines provide  powerful diagnostics of excitation conditions
in   a  stellar   jet  \citep[see   e.g.][]{Bacciotti99,  Hartigan94}:
[\ion{S}{ii}]6716/6731  is  a decreasing  function  of the  electronic
density (until it reaches  a ``high-density limit'' $\simeq 0.45$ when
$n_{\rm  e}\gg  10^4\ {\rm  cm}^{-3}$,  the  critical  density of  the
[\ion{S}{ii}]6716  line); [\ion{N}{ii}]/[\ion{O}{i}] is  an increasing
function    of   the    electronic   fraction    $f_{\rm    e}$;   and
[\ion{S}{ii}]/[\ion{O}{i}]       decreases       with       increasing
temperature\footnote{The   latter  two   ratios  also   decrease  with
increasing $n_{\rm  e}$ when $n_{\rm e}$ exceeds  the critical density
of the line in the numerator}. Hence, these line ratios provide a test
of the heating/ionization  mechanism in the flow, though  they do {\it
not} test the underlying dynamical solution.

\begin{figure}
\begin{center}
\resizebox{6cm}{!}{\includegraphics{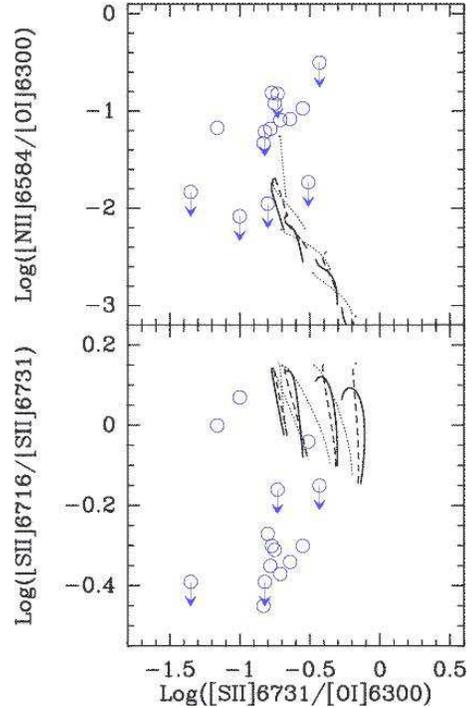}}
\end{center}
\caption{Line ratio diagrams for Model A (solid),  B
(dashed)   and   C  (dotted).    Each   curve   refers   to  a   given
$\dot{M}_{\text{acc}}$, and connects  line ratios at various distances
along the jet, from $z=-50\  {\rm AU}$ to 200~AU (after convolution by
a 70~AU beam  and summation perpendicular to the  jet axis).  Distance
increases       with       [\ion{S}{ii}]6716/6731       and       with
[\ion{N}{ii}]/[\ion{O}{i}] (bottom to top in both figures).  Accretion
rates  vary  from  $\dot{M}_{\text{acc}}=10^{-8}$  to  $10^{-5}\  {\rm
M}_{\odot} \text{yr}^{-1}$ in factors of 10 from left to right in both
figures.}
\label{fig:oasis} %
\end{figure}

In  particular,   \citetalias{Lavalley2000}  and  \cite{Lavalley2000b}
showed how diagnostic  diagrams based on these line  ratios (see their
Fig.~3) could be used to separate various proposed heating mechanisms,
such as shocks, mixing-layers, and ambipolar diffusion heated MHD disk
winds. For the latter case, \citetalias{Lavalley2000} used preliminary
results  from the present  work. We  plot in  Fig.~\ref{fig:oasis} our
complete set of results using the same line ratio diagrams.

Each  curve in  Fig.~\ref{fig:oasis},  corresponding to  a given  wind
model and mass accretion rate, shows the evolution of line ratios with
distance from the  star from 0~AU to 200~AU,  after convolution of jet
images by  a 70~AU  beam and summation  perpendicular to the  jet axis
(note that \citetalias{Lavalley2000} plotted  ratios out to a distance
of 600  AU, for comparison  with observations of  \object{DG~Tau}, and
used  calculations  with  no  depletion  on dust,  which  affects  the
[\ion{S}{ii}]/[\ion{O}{i}] ratio).

Spatially resolved line ratios within 200~AU of the star are available
for only  four TTS  microjets so far:  \object{HH~30}, \object{Th~28},
\object{DG~Tau}, and \object{RW~Aur} \citep{Bacciotti99, Lavalley2000,
Bacciotti2000a,  Bacciotti2000b, Dougados2000b}.  Comparison  with our
predictions, {\it  at the same  beam resolution and distance  from the
star}, indicates  that our models have slightly  too high temperatures
but  insufficient  beam-averaged  electronic  density  and  fractional
ionization.    In  particular,  the   observed  [\ion{S}{ii}]6716/6731
towards the  star is always  in the high-density limit  $\simeq 0.45$,
indicating  $n_{\rm  e}\ge 2\times  10^4\  {\rm  cm}^{-3}$, while  our
models predict higher ratios  at the stellar position corresponding to
a    beam-averaged   $n_{\rm    e}\simeq   0.6-2\times    10^3\   {\rm
cm}^{-3}$. This difference of at least a factor 10-30 readily explains
a substantial fraction of the flux deficit noted above in Section~2.4.
Two further important remarks are in order:

(1) We  stress that,  because  of volume  effects  and strong  density
gradients, beam-averaged  line ratios  differ greatly from  local line
ratios.   In particular  they give  very  little weight  to the  inner
densest parts  of the wind, where the  [\ion{S}{ii}]6716/6731 ratio is
locally in  the high-density  limit.  Hence it  is essential  to apply
proper  convolution  by  the  observing  beam  to  perform  meaningful
comparison with observed ratios.

(2) All  predicted line  ratios, in  particular the  highly ionization
sensitive  [\ion{N}{ii}]/[\ion{O}{i}], would  agree  much better  with
observations  if one  includes extra  mechanical energy  deposition in
shocks,   as  demonstrated  in   \citetalias{Lavalley2000}  \citep[see
also][]{Dougados2000b,  Lavalley2000b},  although  this would  involve
additional free parameters (e.g.  ratio  of shock speed to wind speed,
magnetic field limiting the postshock compression).

For  the latter  reason,  as noted  by  \cite{Bacciotti99}, the  total
density  $n_{\rm  H}  =n_{\rm  e}/f_{\rm  e}$ is  a  more  fundamental
quantity  that can  be directly  compared with  MHD  model predictions
independently  of the  ionization  process (although  $n_{\rm H}$  can
still be affected by shock  compression and by beam averaging). Out of
the  4  above-mentioned T~Tauri  stars  with  spatially resolved  line
ratios, only two  (\object{DG~Tau} and \object{RW~Aur}) have estimates
of disk accretion rates, both  of order $10^{-6}\ {\rm M}_{\odot} {\rm
yr}^{-1}$ \citepalias{Hartiganetal95}.  Observed values of $n_{\rm H}$
as a function  of distance \citep[LF2000;][]{Dougados2000b} are 10-100
times higher than total densities  on the innermost streamline for our
models  with the  same $\dot{M}_{\rm  acc}$ (see  Fig.~1 of  Paper I).
While   shock  compression   could  account   for  a   factor   of  10
\cite{Hartigan94}, it  seems that  our models may  still have  too low
density close to the star.

As  a word  of  caution, we  note that  the  above two  stars are  the
strongest  forbidden  line  emitters  in  the  HEG95  sample  and  may
therefore  not be  representative of  wind physical  conditions  in an
average  TTS. More  sensitive observations  would be  needed  to probe
$n_{\rm e}$  and total density $n_{\rm  H}$ in objects  of weaker wind
emission and lower accretion rates.

\section{Summary and Conclusion}

In a companion paper \citetalias{Garcia2001a}, we computed the thermal
and  ionization structures  of self-similar,  magnetically-driven disk
winds heated by ambipolar diffusion.  In this paper, we confronted our
model  predictions   in  forbidden  lines  with   {\em  all}  existing
observational  constraints  in  T~Tauri  microjets,  namely:  (1)  jet
images;  (2) jet  widths; (3)  long-slit spectra  and  integrated line
profiles; (4)  line fluxes; and  (5) line ratios and  total densities.
Jet widths, line profiles,  and total densities provide powerful tests
of the  jet dynamics and  collimation, whereas line ratios  and fluxes
are strongly influenced by  the underlying heating assumptions.  Given
the  steep   gradients  in  physical  conditions  in   the  jet,  beam
convolution is  found to have  a crucial effect on  model predictions,
and we stress that it is an essential step for a meaningful comparison
with observations.

The  model reproduces  several  observed trends:  (1)  Images show  an
unresolved peak and an extended high-velocity jet, of width compatible
with observations; (2) Line  profiles present a low-velocity component
(LVC), compact and near the star, and a high-velocity component (HVC),
tracing the  jet. These components  are not an inclination  effect but
intrinsic to the jet  dynamics. (3) The systematic differences between
[\ion{O}{i}]  and  [\ion{N}{ii}]  profiles  are reproduced;  (4)  Line
fluxes  are   proportional  to   $\dot{M}_{\rm  acc}$;  and   (5)  the
[\ion{O}{i}] - [\ion{S}{ii}] correlation slope is recovered.

The  model  also encounters  several  problems:  (1)  The line  ratios
([\ion{S}{ii}]6716,6731,  [\ion{N}{ii}]/[\ion{O}{i}])   are  not  well
reproduced,  and indicate  insufficient ionization  in the  model; (2)
Line fluxes are  too low by a factor of  100, and electronic densities
by  a factor  of at  least 10-30;  (3) The  relative intensity  of the
low-velocity component  (LVC), present in  our profiles and  formed in
the inner wind regions just  above the disk, is insufficient; (4) Wind
terminal velocities appear  too high, and total densities  at the wind
base  appear too  small, when  compared  with current  estimates in  2
bright microjets.

Both line intensity  and line ratio problems can  be readily solved by
introducing   an   extra   ionizing   mechanism,   e.g.    in   shocks
\citep{Lavalley2000,    Dougados2000b}.    However,    terminal   wind
velocities  and   total  densities  would  not  be   affected.   As  a
consequence, cold  self-similar MHD  disk wind models  with isothermal
magnetic  surfaces seem  to be  ruled out  by observations  of T~Tauri
microjets.   In the  context of  magnetized  accretion-ejection models
considered  here,  a  possible  improvement  would  be  to  relax  the
assumption  of isothermal magnetic  surfaces.  Indeed,  very efficient
ejection (namely $\xi$ up to 0.5) can be attained by allowing for some
entropy injection  at the disk surface  \cite{Casse2000b}. Such models
are  very  promising  for  they  naturally  provide  both  higher  jet
densities   and   lower   terminal   velocities   (see   Figure~8   in
\cite{Casse2000b}).   In   this  respect,  we   independently  support
\cite{Kwan97}'s point  of view that  active disk chromospheres  may be
necessary ingredients in T~Tauri circumstellar disks.

\begin{acknowledgements}
  
  PJVG  acknowledges  financial  support  from Funda\c{c}\~ao  para  a
  Ci\^encia  e   Tecnologia  by  the   PRAXIS  XXI/BD/5780/95,  PRAXIS
  XXI/BPD/20179/99  grants.   The work  of  LB  was  supported by  the
  CONACyT grant 32139-E.   We thank the referee, Pedro  N. Safier, for
  his helpful comments. We  also acknowledge fruitful discussions with
  Francesca  Bacciotti,  Catherine  Dougados, Pierre  Ferruit,  Eliana
  Pinho, Alex Raga, and Eric  Thi\'ebaut.  PJVG warmly thanks the Airi
  team and his adviser, Renaud Foy, for their constant support.

\end{acknowledgements}

\bibliographystyle{apj}
\bibliography{references}

\end{document}